*Review*

# Why the Brain Knows More than We Do: Non-Conscious Representations and Their Role in the Construction of Conscious Experience

**Birgitta Dresp-Langley**

Centre National de la Recherche Scientifique, UMR 5508, Université Montpellier, Montpellier 34095, France; E-Mail: birgitta.dresp-langley@univ-montp2.fr; Tel.: +33-(0)4-67-14-45-33; Fax: +33-(0)4-67-14-45-55

**Abstract:** Scientific studies have shown that non-conscious stimuli and representations influence information processing during conscious experience. In the light of such evidence, questions about potential functional links between non-conscious brain representations and conscious experience arise. This article discusses neural model capable of explaining how statistical learning mechanisms in dedicated resonant circuits could generate specific temporal activity traces of non-conscious representations in the brain. How reentrant signaling, top-down matching, and statistical coincidence of such activity traces may lead to the progressive consolidation of temporal patterns that constitute the neural signatures of conscious experience in networks extending across large distances beyond functionally specialized brain regions is then explained.

## 1. Introduction

During early childhood, our brain learns to perceive and represent the physical world. Such knowledge is generated progressively over the first years of our lives and a long time before we become phenomenally conscious of the Self and its immediate or distant environment [1]. Statistical learning, or the implicit learning of statistical regularities in sensory input, is probably the first way through which humans and animals acquire knowledge of physical reality and the structure of continuous sensory environments. This form of non-conscious learning operates across domains, across time and space, and across species, and it is present at birth when newborns are exposed to and tested with speech stream inputs [2]. Conscious experience kicks in much later in life, involving complex knowledge representations that support conscious thinking and abstract reasoning [3–5]. How is information represented and processed in the brain to enable such experience? To be able to answer this question, we need to understand how structured knowledge can be represented in neural circuits.

Brain representations have been conceptually divided [6–8] into functionally segregated conscious and non-conscious worlds generating different forms of cognition and awareness. How the different cognitive worlds interact to produce successful adaptive behavior at the least possible cost is not known, but a large number of studies have shown that non-conscious brain processes influence perceptions and representations embedded in ongoing conscious experience. Non-conscious brain processes have the capacity of encoding vast amounts of information relative to complex events of the physical world through multiple interdependent sensory channels at any given moment of time. Conscious processing, on the other hand, is extremely limited in capacity, which explains why most of our knowledge of the world is generated outside consciousness [9]. Whenever we consciously remember, decide, or act, our brain seems to "know" far more about what we are doing and why we are doing it than our conscious experience is able to consider. It appears that, as a result of evolutionary pressure and selection, the human brain has achieved to govern complexity at the least possible cost by selectively allocating resources, at the level of our sensations, emotions, decisions, and actions.

Theoretical models have tried to explain how such a selective process may work by suggesting that non-conscious sensorial and representational processes interact, through working and long-term memory, to generate brain learning and, ultimately, enable conscious experience [10–13]. Some of these have defended the idea that non-conscious signals and memory representations are selectively made available to conscious experience on the basis of temporal coincidence of representations at a given moment in time. This would involve neural mechanisms that match distributed signals from non-conscious and conscious levels of brain processing into time dependent representations of knowledge and events. On the grounds of functional hypotheses of these and earlier theories [14–16], it is possible to clarify some major functional implications of non-conscious brain representation for the generation of conscious experience: (1) Only non-conscious brain processes have enough capacity to process the complex cross-talk between signals originating from various simultaneously activated and functionally specific sensory areas; (2) The temporal signatures of conscious experience are formed and consolidated in reverberating interconnected neural circuits that extend across long distances and well beyond functionally specific brain regions; (3) This is achieved though the matching of coincident neural activity traces of non-conscious memory representations; (4) The temporal signatures of conscious experience are independent from spatial brain maps and remain available after destruction of the specific functional circuits through which they have been originally formed.

The following section presents evidence which supports the general idea defended here that non-conscious brain representation enables a selective process for either making representations accessible to, or suppressing them from, immediate consciousness. Then, Section 3 introduces some of the major functional assumptions of a brain model capable of explaining how such a selective process could work, and Section 4 clarifies how temporal neural signatures of conscious experience could be generated on the basis of non-conscious brain processing. In Section 5, a conclusion and further perspectives are given.

## 2. Non-Conscious Perception Influencing Conscious Cognition and Action

Non-conscious processes in emotional, perceptual and cognitive function cover a wide range of observations, such as subliminal psychodynamic activation and non-conscious perception in hypnosis, subliminal semantic priming, effects of stimuli that are not consciously perceived on recognition processes, implicit learning, the perceptual integration of subliminal luminance or color targets , and phenomena of blind-sight in patients with striate cortical lesions, non-human primates, and normal observers Subliminal psychodynamic activation or *SPA* [17] describes behavioral effects where the exposure to subliminally presented, drive-related stimuli results in a positive change in the emotional and mental state of human observers [17,18]. So-called "symbiotic" imagination or fantasies, where comforting internal representations are triggered by comforting subliminal stimulation, for example, are key issues here. Results from clinical studies have shown that subliminal verbal messages designed to induce such symbiotic fantasies and administered under double-blind quasi-experimental conditions significantly reduce anxiety levels and raise the motivation of psychiatric patients such as drug abusers [19]. Follow-up examinations furthermore revealed that the experimental patient groups who received treatment with the subliminal stimuli reported more dreams containing positive symbiotic events than the controls. It is emphasized that the non-conscious character of the stimuli in subliminal psychodynamic activation (SPA) is critical: effects produced under conditions where observers are unaware of the nature and content of the stimuli were found to be significantly stronger than those produced by the same stimuli presented at supraliminal levels [20]. Explanatory models of SPA effects suggest that supraliminal stimuli lose some of their power to produce the desired effects on internal representations because subjects perceive them as part of an externally administered procedure [21]. In other words, stimulus awareness would in this case diminish the organisms' capacity for responding efficiently to drive- and affect-related stimuli. Some restricting effect of awareness on psychodynamic responsiveness is widely believed to diminish the efficiency of relaxation techniques that combine soothing music with verbal suggestions, which has lead to the sustained use of subliminal suggestions combined with soft music in relaxation therapy. Experimental studies have shown that the most efficient combinations appear to be indeed those where soft music is presented together with verbal stimuli of intensities below the level of conscious perception [22].

Theory and findings regarding SPA effects have received critical feed-back raising issues relating to the appropriateness of control and threshold stimuli in the various experiments [23,24], the possible need for physiological indicators of anxiety reduction such as the subject's heart rate in addition to the psychological measures [25], and questions about the need for neutral, *i.e.*, neither drive- nor affect-related, stimuli to establish individual subjective thresholds for SPA [26]. However, quantitative and qualitative reviews and meta-analyses of research conducted over the years lead to the conclusion that the major findings remain statistically significant [27]. Partial-cue hypotheses, which suggest that some structural cues in subliminal stimuli might be directly available to consciousness and therefore explain SPA effects, have been put into question [20].The implications of the initial observations for cognitive science remain the same: conscious processing interferes with responding optimally to drive- and affect-related stimuli.

Hypnosis and hypnotic suggestibility are phenomena that are not yet fully understood scientifically, but merit nonetheless attention. Hypnotic induction is not as such a subliminal process since the psychodynamic effects in this case are mediated via essentially supraliminal verbal suggestions. Hypnotic phenomena may reflect states of altered consciousness [28], and the degree to which a human individual may respond to hypnotic suggestions is referred to as hypnotic susceptibility, which can be accurately predicted on the basis of psychometric tests such as the Waterloo-Stanford Group Scale of Hypnotic Susceptibility [29,30]. Hypnotic susceptibility is an estimate of the ability of a man or a woman to enter some trance-like state where overall awareness is shifted away from the general context and environment, and focused on the symbiotic fantasies induced by the hypnotic (verbal) suggestions of an expert clinician. Hypnotic suggestibility in young men and women has been shown to be significantly enhanced following application of weak (one micro Tesla) burst-firing magnetic fields for 20 min over the temporal-parietal lobes of the right hemisphere [31]. This suggests that the signatures of the low-frequency magnetic fields contain bio-relevant information which directly affects the neural processes underlying hypnotizability. Positron emission tomography (PET) measures of regional cerebral blood flow and electroencephalographic (EEG) measures of brain electrical activity have shown that specific patterns of cerebral activation are associated with the hypnotic state and the processing of hypnotic suggestions [32]. Another PET study comparing highly susceptible males with an additional ability to hallucinate under hypnosis, so-called "hallucinators", to other highly hypnotizable "non-hallucinators" revealed that a specific region in Brodman area 32 was activated in the group of "hallucinators" when they heard an auditory stimulus, or when they merely hallucinated hearing it under hypnosis [33]. Such activation was absent when the "hallucinators" merely imagined hearing the tone, and in all "non-hallucinators" regardless of experimental condition.

Measurable consequences of hypnosis intervention on cognitive function were also reported. With highly susceptible observers, hypnosis may produce an inhibition of correct responses in perceptual tasks with conflicting stimuli [34], correlated with significant changes in cortical evoked potentials. Effects of hypnotic susceptibility on auditory event-related potentials (AERPs) were found in observers who were instructed to ignore tones while accomplishing some other task, such as reading a novel. The highly hypnotizable subjects revealed different AERP amplitudes and latencies when ignoring the tones, and were significantly slower in responding to the not-to-be-attended stimuli compared with less susceptible subjects. This suggests that highly hypnotizable humans may have a greater ability to shift awareness towards relevant stimuli and away from irrelevant ones [35]. Furthermore, specific hypnosis techniques such as suggested selective deafness or selective visualization appear to influence learning processes in the desired direction, which means that subjects under hypnosis are able to eliminate from consciousness exactly what they are told to [36].

A particular example showing how supraliminal perceptions or representations may become subliminal through guided shifts in awareness induced by hypnotic suggestions is the hypnotic control of physical pain, or hypnotic analgesia [37,38]. Research on hypnotic analgesia has grown substantially in recent years and helped to develop strategies for acute and chronic pain management in the private and public domains. Although it is often difficult to distinguish facts from *artifacts* such as placebo, current knowledge points towards the general agreement that pain and distress perception is significantly lowered through hypnosis, in acute as well as chronic pain patients with high hypnotic susceptibility [39]. Recent scientific studies have investigated the effect of hypnotically induced obstructive fantasies, targeted at eliminating painful stimuli from consciousness. The results showed significantly higher pain and distress tolerance, significant changes in EEG amplitude, and a significantly reduced heart rate [40] in highly susceptible individuals subjected to painful electrical stimulation under hypnosis. Target specific amplitude peaks in response to somatosensory stimuli were found significantly reduced in subjects with high hypnotizability in a pain target detection task [41]. Apart from the possible implications for clinical research, these effects of hypnosis suggest that perceptions and representations embedded in ongoing awareness can be selectively eliminated from consciousness [38].

Effects where a person's conscious feelings, judgment, or choices are changed by non-conscious processing of images or messages have also been reported. Such is supposed to happen every time when we watch television or look at colorful adverts in a magazine or in the street [42]. In a BBC broadcast study by Underwood [43], faces were flashed subliminally within the program for about 20 ms in a restricted part of the network region. Immediately after the broadcast, TV viewers were invited to make a judgment by telephone about a neutral, supraliminal face image that expressed no emotion. Judgments were made by telephoning one of two numbers (1 or 2) indicating "sadness" or "happiness". Statistical analyses of the phone call responses revealed that viewers who received a subliminal smiling face in the broadcast were less likely to judge the neutral face as being happy than were those viewers who were not exposed to the subliminal image in the program. Underwood suggested that this effect could be explained in terms of a contrast effect, where the neutral expression of the supraliminal image is interpreted as "sadder" than the smiling subliminal image. However, the broadcast study provided no information as to whether the so-called subliminal frames may have been perceived in some cases. Attempts to replicate the results of the broadcast study under laboratory conditions did not yield findings unambiguous enough to allow for a clear conclusion. Emotional priming is often difficult to control and depends on a variety of factors, ranging from the graphic quality of the material presented to the general mood of a given individual subject at a given moment. It can therefore be expected that, when priming people with emotions, both contrast and assimilation effects may occur. Also, some stimuli may have a particular status in subliminal emotional priming [44] given that differences in the detection thresholds of different subliminal images or stimuli were reported, with the lowest thresholds for the subjects' own name and images of happy faces. Results from other studies [45] suggest that subliminally presented pictures of angry faces may yield stronger emotional responses compared with images of non-consciously perceived happy faces, especially in men. Some of the apparent inconsistencies in results from experiments designed to influence emotional responses through non-conscious stimuli have fed the feathers of those eager to dismiss such evidence altogether. However, we must bear in mind that emotional cognition depends on a variety of epigenetic variables, such as personal experience and culture, and that results can be expected to vary considerably as a function of the latter.

Research on the influence of non-conscious perception on cognitive processes such as recognition, memory, and learning harks back to the experimental studies by Marcel, who investigated effects of visual masking on word recognition. These early findings [46] showed that conscious processing of visual objects is not necessary for their subsequent recognition, motivating further studies on subliminal semantic priming, where clearly visible targets are preceded by non-perceptible stimuli, so-called primes. Non-conscious primes have been found to directly influence the conscious processing of supraliminal targets. For example, near-threshold visual primes in a memory task significantly increases the recall of items that are not recalled when presented without the primes, despite the fact that the reported "feeling of knowing" of observers did not change between the two experimental conditions [47].

Non-conscious processing therefore effectively supports conscious representation without individuals being aware of the immediate behavioral outcome. In experiments where subjects had to classify visually presented words (targets) into semantic categories, significant effects of semantically congruent non-conscious primes, producing significantly lower error rates, were reported [48]. The prime words were rendered undetectable through masking and brief exposure durations between 17 and 50 ms, and observers were instructed to respond within a narrowly restricted time window. The magnitude of priming effects as a function of prime visibility [49] has been investigated using linear regression analysis, showing that conscious semantic representation is particularly facilitated by non-conscious primes, suggested by the results of a multitude of studies on memory without awareness [50,51].

Neurophysiological studies have provided insight into the brain correlates of semantic priming, using a combination of behavioral task and brain-imaging technique [51]. It was shown that non-conscious prime stimuli have a measurable influence on the electrical and hemo-dynamic characteristics of brain activity. Other functional neuro-imaging studies have investigated brain correlates of the so-called "mere exposure effect" [52–54]. The latter describes observations where mere pre-exposure to visual objects that are not identifiable beyond the chance level is sufficient to significantly influence subsequent preference and memory recall of consciously perceived objects. The "mere exposure effect" may thus be seen as a variation of subliminal semantic priming since it suggests, like priming effects in word recognition, that non-conscious processing impacts on conscious memory judgments. In groups of subjects making memory and preference judgments about consciously perceived objects after previous exposure to subliminal stimuli, specific neural activities in the right lateral prefrontal cortex associated with the implicit memory retrieval process were identified [55]. The data appear consistent with earlier evidence for right lateral prefrontal activation during implicit behavioral guidance without awareness [56], and have been interpreted in terms of a particular memory system operating outside consciousness. Subjects were not aware that they had been exposed to their preferred or correctly recalled objects before, whereas their brains had processed the subliminal information effectively. Associative learning without conscious report has been tagged by specific temporal patterns of event-related potentials (ERP). ERP activities triggered by aversive responses (shock-*versus*-no-shock aversive conditioning) to non-consciously perceived faces were compared to activities triggered by aversive responses to consciously processed faces [57]. Specific temporal activity patterns indexing the acquisition of a conditional response to the non-consciously processed faces were found, supporting the idea that brain traces of classical conditioning are formed in circuits that control non-conscious as well as conscious behavioral processes or experience. Visual perceptual learning experiments have shown that subliminal presentation of one of two contingent signals in a choice reaction time task yields the same learning performance as presentation of two consciously perceived contingent signals while subjects were unable to recall the nature of the non-consciously induced contingencies after learning [58]. Similarly, non-consciously induced auditory affirmations embedded in soft background music have been found to influence the learning and conscious recall of semantically ordered lists [59]. Subjects who were exposed to subliminal voice input did significantly better in recalling items from the lists than controls.

Evaluative conditioning is a particular case of emotional priming. Changes in emotional responses to stimuli that are supposed to be affectively neutral at the beginning can be primed in either a positive or negative direction, as in the controversial experiments by Underwood mentioned earlier here, by introducing a subliminal associative stimulus. After repeated association with stimuli carrying strongly negative or positive emotional connotations, the initially neutral stimuli then elicit emotionally biased responses. Some studies using evaluative conditioning have shown that the conscious evaluation of objects judged "neutral" at the beginning changed towards "negative" or "positive" judgments after a series of trials where the presumed neutral objects have been associated repeatedly with either a positive or negative, non-consciously perceived stimulus [60]. Such observations have been interpreted in terms of subliminal contingency learning through interactions between non-conscious emotional responses and conscious decisions about "good" and "bad", or values and norms in general. Whether or not non-consciously perceived emotional stimuli suffice to produce reliable, firmly consolidated contingency learning has been put into question by results from more recent experiments. For example, item-based analyses of responses to individual stimuli from several experiments have led to the conclusion that consistent evaluative conditioning only emerges when, in the course of the learning or valence acquisition process, subjects are made aware of the nature of the contingency between a neutral stimulus and an emotionally biased, positive or negative, associated stimulus [61]. Weaker effects from previous studies with merely subliminal contingencies [60] may partly be explained by factors such as inter-individual differences in attention or readiness to respond. However, valence acquisition through repeated contingency priming is a rather particular learning process, where initially undetermined or ambiguous emotional representations can shift towards strongly biased ones in either of two strictly opposite directions. To achieve stable output from such learning may require conscious control at critical moments, and the more recently reported necessity of momentary contingency awareness [61] may reflect a mechanism that fulfills an important functional role in the consolidation process, as will be explained in the following chapter in the light of the model proposed here.

Accounting for the effects of non-conscious sensory stimuli on conscious behavior requires making a clear distinction between the sensory threshold, that is the psychophysical or statistical threshold for the detection of a stimulus as defined by Signal Detection Theory [62], and other thresholds for the semantic processing of stimuli, such as recognition or identification thresholds. A subliminal sensory signal or stimulus is defined as a signal with intensity levels below the psychophysical detection threshold. During exposure to a psychophysically subliminal stimulus in a visual task, a human observer may sometimes be aware of the fact that he/she may have seen something, but will not be able to say what it was, or be unaware of other specific characteristics. The influence of subliminal signals on the spatial and temporal integration of contrast stimuli has been investigated psychophysically for a long time [63,64], showing that non-conscious signals influence conscious vision. Electrophysiological studies have shown significant event-related brain responses to subliminal visual stimuli [65], where a specific signal component could be assigned to the processing of a non-conscious visual target.

Evidence for a shift from non-conscious to conscious sensory processing as a function of visual context has been found in psychophysical experiments with supraliminal contrast lines and subliminal contrast targets collinear with the lines. In these experiments, small vertical contrast lines (targets) had to be detected at fixed locations. In some conditions, the targets were presented together with a clearly visible, spatially separated collinear line (context); in others, the targets were presented alone (no context). While the contrast intensities of these targets were mostly subliminal, or non-detected without the context, they became detectable when presented with the context [66–68]. Subliminal color targets were also found to become detectable when presented together with consciously perceived colored lines or edges, but needed longer exposure durations for the effect to occur compared with achromatic versions of the same stimuli [69,70]. Brain correlates of this phenomenon have been identified in the visual cortex of a wakeful behaving monkey accomplishing a similar psychophysical detection task [71], where neural activities triggered by targets were found to be increased by the presence of a clearly visible, spatially separated, collinear line, and diminished by the presence of a perpendicular line. These studies have led to identifying the underlying neural mechanisms in terms of long-range interactions, suggesting that the latter may be involved in generating interactions between conscious and non-conscious visual signals.

However, neural pathways other than those projecting to striate cortex seem to be involved in generating the non-conscious processing of visual signals. In patients with cortical blindness caused by lesions to their primary visual cortex (striate cortex V1), residual responses to visual objects are found while observers are unable to report what they actually see [72]. Such patients accurately detect monochromatic visual stimulus patterns, can discriminate direction of movement as well as orientation of stimuli in their "blind" fields, and are able to discriminate the wavelength of chromatic stimuli in the absence of any consciously acknowledged perception of color [73]. Whether the loss of all conscious vision is an inevitable consequence of striate cortical destruction has remained unclear. Patients with homonymous right hemianopias tested in tasks designed to assess their perception of visual objects presented within the blind field were capable of making appropriate preparatory manual adjustments (reaching and grasping) and seemed able to consciously process structural and semantic characteristics of such objects [74]. Monkeys with unilateral removal of V1 preserve residual visual capacity in the sense that the animals can still detect and localize visual signals in their affected hemifields, but do not seem to be able to identify the nature of these signals [75]. Non-conscious visual processing thus influences conscious action in humans [76] as well as animals.

## 3. The Functional Role of Non-Conscious Representation

Non-conscious brain processes are presumed to have the capacity of processing a majority of incoming signals and to hold them available for further processing, after selection, at the conscious level, which is limited in capacity [10,12,13]. Visual search studies have shown that observers search faster and are more efficient when they are not conscious of what they need to be looking for [77]. It also has been shown that non-conscious information processing is not only faster, but also capable of generating multidimensional knowledge of interactive relations between variables that are too sophisticated to be processed consciously [78]. A great deal of human decision making in everyday life occurs indeed without individuals being fully conscious of what is going on, or what they are actually doing and why. Also, human decisions and actions based on so-called intuition are quite often timely and pertinent and reflect the astonishing ability of the brain to exploit non-conscious representations for conscious action, effortlessly and effectively. Non-conscious representation is aimed at reducing complexity at the level of conscious processing. It enables the brain to select, from all that it has learnt about outer and inner events, only what is needed for producing a meaningful conscious experience.

*3.1. Adaptive Resonance and Brain Learning*

Adaptive resonance theory [14] conceives the brain as a knowledge generating machine with multiple, parallel distributed unit structures. It has given valuable conceptual support for thinking about how different processing levels may produce coherently organized knowledge structures, how context-sensitive adaptive learning may generate non-conscious representations, and how these latter can be made available to conscious experience at a given moment in time, generating meta-representations of knowledge that become embedded in a single conscious experience. In neural networks, cells can become subliminally active when they receive priming signals that sensitize or modulate their actual response or responsiveness by preparing them to react more quickly and vigorously to subsequent bottom-up inputs that match the priming signals [79]. Perceptual knowledge of a visual environment, for example, would require that subliminal mechanisms be present in every cortical area wherein learning can occur, since without such mechanisms, any learned knowledge would be rapidly degraded and subject to what Grossberg [14] refers to as "catastrophic forgetting". Neural network models specifically developed to account for subliminal priming effects [80] suggest modifications of neural reaction times to subsequent inputs, according to whether or not there are traces of subliminal processing of earlier input. Such models use parallel processing modules, or cell assemblies, with different lateral connectivity and output functions. Their functional properties are consistent with the hypothesis that the human brain uses parallel codes for the representation of contents or knowledge, and that these codes generate a conscious state when the discharges of functionally related neurons match in the domain of knowledge and in the domain of time. Non-conscious brain mechanisms would serve the purpose of boosting relevant bottom-up signals and suppressing irrelevant signals at the appropriate time, and thus lead to a constant updating of current representational knowledge outside consciousness. Temporal summation at dendrites of hippocampal neurons in the rat [81], obtained with a technique where the strengths, sites, and timing of dendritic inputs can be controlled with precision, reveal that the temporal integration of synaptic inputs can readily switch between subthreshold and suprathreshold summation. This seems to suggest that active conductance in concert with passive cable properties in biological neural networks may serve to boost coincident synaptic inputs and to attenuate or suppress non-coincident inputs. Such properties of synaptic transmission could be exploited by brain mechanisms to generate interactions between specific, temporally related subliminal and supraliminal signals.

Earlier cognitive theories had suggested that perceptions and sensations may feed into functionally separate processing streams, operating within or outside consciousness [7,82]. Kihlstrom [6], in particular, suggested that conscious processing is functionally dissociated from perceptive-cognitive functions such as discriminative responses to sensory input, perceptual skills, memory, and higher mental processes involved in decision making, judgment and problem solving. On this basis, he proposed taxonomies for what he referred to as "the cognitive unconscious". Kihlstrom emphasized that humans seem to be able to perform cognitive analyses on information which is not itself accessible to awareness by means of automatic and unconscious procedural knowledge. He suggested a tripartite division of the "cognitive unconscious" into "truly unconscious", "preconscious", and "subconscious" parallel processes. These three would run in parallel with a "truly conscious" processing stream that generates declarative knowledge structures. Kihlstrom's theory suggests four parallel processes to account for the ways in which the brain generates knowledge. Mechanisms that would explain how the brain passes from one level to another are not suggested.

The idea of a strict functional segregation between the cognitive unconscious and conscious experience may have to be reconsidered. The brain appears to process information through circuits which interact at multiple levels of integration and across large distances in the brain, well beyond intrinsic functional specialization [83–87]. It seems plausible to suggest that subliminal perceptual input is processed and represented in all areas of the brain capable of generating resonant interactions, where subliminal representations are made temporally available to conscious experience on the basis of mechanisms detecting coincident representations. Representations embedded in conscious experience can then also be temporally suppressed on the basis of these same mechanisms. One of the problems with such a conceptual framework consists of explaining how subliminal input traces can be processed and stored in neural structures without interfering with ongoing processing or, more importantly, without destroying or changing representations that are already stored.

*3.2. Non-Conscious Representation Matching*

Attempting to solve this problem, Grossberg [14] defined specific functional principles for the generation of non-conscious representations in resonant circuits of the brain. These functional principles exploit two mechanisms of neural information processing, referred to as bottom-up automatic activation and top-down matching.

(1) Bottom-Up Automatic Activation is a mechanism for the processing and the temporary storage of perceptual input outside conscious experience. Through Bottom-Up Automatic Activation, a group of cells within a given neural structure becomes supraliminally active whenever it receives the necessary bottom-up signals. These bottom-up signals may or may not be consciously experienced. They are then multiplied by adaptive weights that represent long-term memory traces and influence the activation of cells at a higher processing level. Grossberg [14] originally proposed Bottom-Up Automatic Activation to account for the way in which pre-attentive processes generate learning in the absence of top-down attention or expectation. It appears that this mechanism is equally well suited to explain how subliminal signals may trigger supraliminal neural activities in the absence of phenomenal awareness. Bottom-Up Automatic Activation generating supraliminal brain signals, or representations with adaptive weights near or at zero, would be a candidate mechanism to explain how the brain manages to process perceptual input that is either not relevant at a given moment in time, or cannot be made available to conscious processing because of a lesion in the circuitry, as in vision without consciousness for example.

(2) Top-down Representation Matching is a mechanism for selectively matching bottom-up representations of incoming signals to learnt memory representations. Subliminal representations may become supraliminal when bottom-up signals or representations are sufficiently relevant at a given moment in time to activate statistically significant matching signals. These would then temporally match the bottom-up representations (coincidence). A positive match confirms and amplifies ongoing bottom-up representation, whereas a negative match tends to invalidate ongoing bottom-up representation [14]. Top-down matching thus may be conceived as a selective process where non-conscious representations become either embedded in, or remain temporarily inaccessible to, conscious experience.

The matching rules address the so-called attention-pre-attention interface problem [14] by allowing pre-attentive (bottom-up) processes to use some of the same circuitry that is used by attentive (top-down) processes. This would help stabilize cortical development and learning. Top-down matching in its most general sense generates feed-back resonances between bottom-up and top-down signals to rapidly integrate brain representations and hold them available for a consciousness experience at a given moment in time. Non-conscious semantic priming can be explained on these grounds. Statistically significant positive top-down matching signals produced on the basis of strong signal coincidences would explain why subliminal visual representations become consciously perceived when presented simultaneously with a specific context, especially after a certain amount of visual learning or practice [88]. Conversely, significant negative matches produced on the basis of repeated discrepancies generating strong negative coincidence signals could explain why a current conscious representation is suppressed and replaced by a new one when a neutral conscious representation is progressively and consistently weakened by association with a strongly biased representation, as in evaluative conditioning and contingency learning [60,61]. The above mentioned functional properties require long-range connectivity of cortical circuits capable of generating what Edelman [4] called "reentrant signaling". Bottom-up representations activating specific structures of such circuits, but not producing sufficiently strong matches to long-term memory signals, will remain non-conscious. Strong positive top-down matching of selected representations will compete with weaker or negative matches and, ultimately, produce their suppression from an ongoing conscious experience, as for example in psychodynamic suppression, where sudden conscious integration of new input interferes with the ongoing conscious processing of other stimuli. Also, specific instructions telling subjects what to expect or what to attend to can thereby generate top-down expectation signals strong enough to inhibit matching of other relevant signals at the same moment in time, as for example in the hypnotic control of physical pain. Strong negative top-down matching reflects a competing process, generating output of the opposite sign, *i.e.*, a negative instead of positive coincidence index. Results from certain observations in motor behavior, which operates mostly outside awareness [89], highlight potential implications of negative top-down matching for conscious control in learning processes with conflicting intermediate output. For example, it has been shown that individuals may become aware of unconsciously pursued goals of a motor performance or action when the latter does not progress well, or fails [90]. This could reflect the consequence of repeated negative top-down matching of the non-conscious bottom-up goal representation and top-down expectation signals in terms of either memory traces of previous success, or representations of desired outcome. Repeated and sufficiently strong negative matching signals might thereby trigger instant awareness of important discrepancies between expectancy and reality. In contingency learning where an expectedly neutral stimulus is repeatedly associated with a biased one, conscious control might be necessary to reinforce and consolidate new representations of the neutral stimuli. Subliminal exposure to a biased associative stimulus without contingency awareness might fail to produce such consolidation because the negative matching signals may in this case not be strong enough to outweigh the old representation. It is, indeed, likely that conscious control in any learning under conditions of high uncertainty fulfills an important adaptive function that has evolved in response to pressure from a steadily changing environment.

## 4. How Does the Brain Link Non-Conscious Representations to Generate Conscious Experience?

Dresp-Langley and Durup [13] suggested that non-conscious representations are linked to conscious experience through coincidences of neural activity patterns in resonant brain circuits. The term representation is defined here as by Churchland [91] in terms of patterns of activity across groups of neurons which carry information. Such patterns of neural activity are described by signals distributed across time and forming unique sequences. They constitute the potential temporal signatures of conscious experience.

*4.1. The Temporal Signatures of Conscious Experience*

Several approaches have suggested functional properties to explain how groups of neurons could produce specific temporal signatures through timing-dependent mechanisms where bottom-up processing is represented by the temporal firing activity of a specific coding assembly for a specific temporal window or duration. The activity traces of long-term memory representations would then consist of unique combinations of many such temporal sequences [15], generated within reentrant circuits of neurons with widely extending long-range connections. Dresp-Langley and Durup [13] proposed that, whenever such memory traces generate significant reentrant matching signals in the dedicated resonant circuitry, a conscious experience is triggered, and its unique temporal signature is "printed" in the brain. This signature remains potentially available for a new conscious experience, and can be retrieved again through top-down matching.

John [12] suggested that a conscious experience may be identified with a brain state where information is represented by levels of coherence among multiple brain regions, revealed through coherent temporal firing patterns that deviate significantly from random fluctuations. These assumptions are consistent with the idea of stable and perennial temporal signatures of conscious experience. These latter arise from temporal interactions between non-conscious representations and are preserved when spatial remapping or cortical reorganization takes place. Empirical support for this theoretical framework comes from evidence for functional links between electroencephalographic activities and consciousness [92]. A temporal activity index signaling coherent firing patterns (coherence index) was computed, and found to change significantly with increasing sedation in anesthesia, independently of the type of anesthetic [93] Decreasing temporal activities were reported when doses of a given anesthetic were increased. Characteristic temporal activity patterns signaling coherence occur across brain regions during focused arousal and during REM sleep, when the subject is dreaming [94] They disappear in dreamless, deep slow-wave sleep, which is consistent with the findings on deeply anesthetized patients, suggesting that the temporal brain signatures of conscious experience are activated in dreaming, which is consistent with [95], who suggested earlier that dreams and conscious imagination represent equivalent conscious experiences.

The temporal activity matching of non-conscious representations for the generation of conscious experience results from intra-cortical reverberation and may correlate with brain mechanisms which establish arbitrary but non-random departures from different functional regions or topological maps, which may be subject to functional re-organization. Thus, conscious experience is constructed on the basis of selective matching of non-conscious representations. This requires reentrant brain circuitry, long-range inter-cortical connectivity and, most importantly, functional plasticity. Such brain properties should make it possible to retrieve any given temporal signature through any set of coding cells. The neural basis of conscious experience is then identified with specific temporal properties of resonant activity patterns, arbitrarily determined through brain learning.

*4.2. Functional Implications of Long-Distance Reentrant Signaling*

Reverberant circuits or loops in the brain have their own intrinsic functional topology [96,97] and were found in thalamo-cortical [98] as well as in cortico-cortical pathways [99,100]. Reverberant neural activity, or reentrant signaling, is a purely temporal process that generates feed-back loops in the brain. It reflects an important functional property of the brain because without it, the conscious execution of focused action would be difficult, if not impossible [101]. This has led to suggest that consciousness relies on the extension of local brain activation to higher association cortices that are interconnected by long-distance connections forming reverberating neuronal circuits extending across distant perceptual areas. A major functional advantage of such long-distance reverberation would be that it may enable the neural traces of non-conscious representations formed in functionally specific circuits to travel well beyond their functional boundaries. Functional imaging studies have associated conscious brain activity with the parieto-frontal pathways, others suggested occipital correlates [86,87]. What both these brain regions have in common, interestingly, is that they are protected from fluctuations in sensory signals and therefore allow information sharing across a broad variety of higher cognitive processes. We argue that such selective information sharing leads to a significant reduction in bottom-up signal variations, which provides a clear functional advantage for the top-down matching of non-conscious representations at the least possible cost in terms of information processing. Sorting out highly complex cross-talk between signals from a multitude of different sensory channels is then no longer necessary. Moreover, long-distance reverberation of neural activity traces across long-range connections enables the functional segregation of spatial contents from their temporal traces, which clarifies how a stable and precise brain code can be generated despite the brain's highly plastic and largely diffuse spatial functional organization and thereby resolves the *stability versus plasticity* dilemma [14]. A candidate mechanism for explaining how this may work is signal de-correlation, an important concept in neural network theory and systems theory in general. Signal de-correlation reduces cross-talk between multichannel signals in complex systems while preserving other critical signal properties and could therefore be an important aspect of selective brain processing, with an undeniable adaptive advantage to any species having evolved such capacity [13].

*4.3. Cortical Plasticity and Epigenetic Factors*

A multitude of sensory, somatosensory, and proprioceptive signals can be perceived simultaneously in a single conscious moment. The integration of such a variety of signals into a unifying conscious experience originate relies on the temporal linking of non-conscious representations, which have to be stable, yet, possess functional plasticity to enable the continuous updating of representations in response to changes in context. This allows the brain to learn and integrate new facts, events, and event properties. Clinical observations and case studies of the "phantom limb" syndrome [102] are consistent with the idea of a highly plastic functional organization of the brain. The phantom limb syndrome was first mentioned in writings by Paré and Descartes, referred to and described in great detail by [103]. It has been repeatedly observed in hundreds of case studies since. After arm amputation, patients often experience sensations of pain in the limb that is no longer there. Experiments with such patients have shown that about a third of them systematically experiences stimulations of the face as coming from the phantom limb, with a topographically organized map that matches individual fingers of a hand. Similar evidence for massive functional re-organization of somatotopic maps after digit amputations has been reported since. For example, several years after dorsal rhizotomy, a region corresponding to the hand in the cortical somatotopic map of the adult monkey brain can be activated by stimuli delivered to the face [104]. It has been suggested that cortical remapping should be possible everywhere in the higher brain, and massive functional plasticity [105] would explain how brain traces of non-conscious representations remain available to conscious experience, even when the original circuits which have built them are destroyed.

It seems that during brain learning, the progressive selection of coincident activity traces of non-conscious representations builds some kind of access code for conscious experience. This is achieved on the basis of purely statistical criteria and progressively leads to fewer and fewer consolidated patterns for the increasingly complex signal coincidences the brain is to learn throughout its epigenetic development. When we are born, all brain activity is more or less arbitrary, but not necessarily random. During brain development, temporal activity patterns elicited by events in biophysical time will be linked to a variety of particular conscious experiences in a decreasingly arbitrary manner. Frequently activated patterns are progressively consolidated through a process of developmental selection [4,13]. The idea of developmental selection of temporal signatures of conscious experience resolves a critical problem in Helekar's model [15], which fails to explain how the non-arbitrary linking of non-conscious representations could work. Helekar seemed to be aware of this problem and proposed a genetically determined linkage which is, however, inconsistent with the fact that brain learning is experience dependent and, despite universal principles and prewired functions, strongly influenced by epigenetic factors. Helekar's elementary, experience-coding temporal activity patterns are generated by prewired subsets of neural patterns from all patterns the brain could possibly generate. His hypothesis was that only patterns that are members of this prewired subset would be involved in the generation of conscious experience.

I prefer to think that non-conscious representations are encoded and decoded in the brain through mechanisms of neural learning which, although they may well be universal [106], express themselves not through some genetic program, but on the basis of developmental processes which are themselves experience-dependent. Such processes can ensure the non-arbitrary linkage of non-conscious contents and their brain traces. Once such traces are matched to form the temporal signature of a new conscious experience, this signature remains potentially available as a "brain hypothesis". This hypothesis is then is either progressively reinforced and consolidated, or slowly extinguished. Once consolidated, the linkages between non-conscious representations become less arbitrary, in some cases deterministic. Grossberg [11] himself often invoked evolutionary pressure to explain why resonant brain mechanisms make good sense. Let us go one step further and suggest that evolution has produced brains capable of generating conscious experience on the basis of a higher and more abstract level of functional organization than previously imagined, where spatial aspects of information processing are discarded and only the temporal traces of non-conscious representations preserved to be matched for complex conscious experiences at any given moment in time (Figure 1), with or without external stimuli.

**Figure 1.** The statistical selection of matched temporal activity traces of non-conscious brain representations builds the neural signatures of conscious experience in biophysical time. Psychological time associated with a conscious experience is subjective.

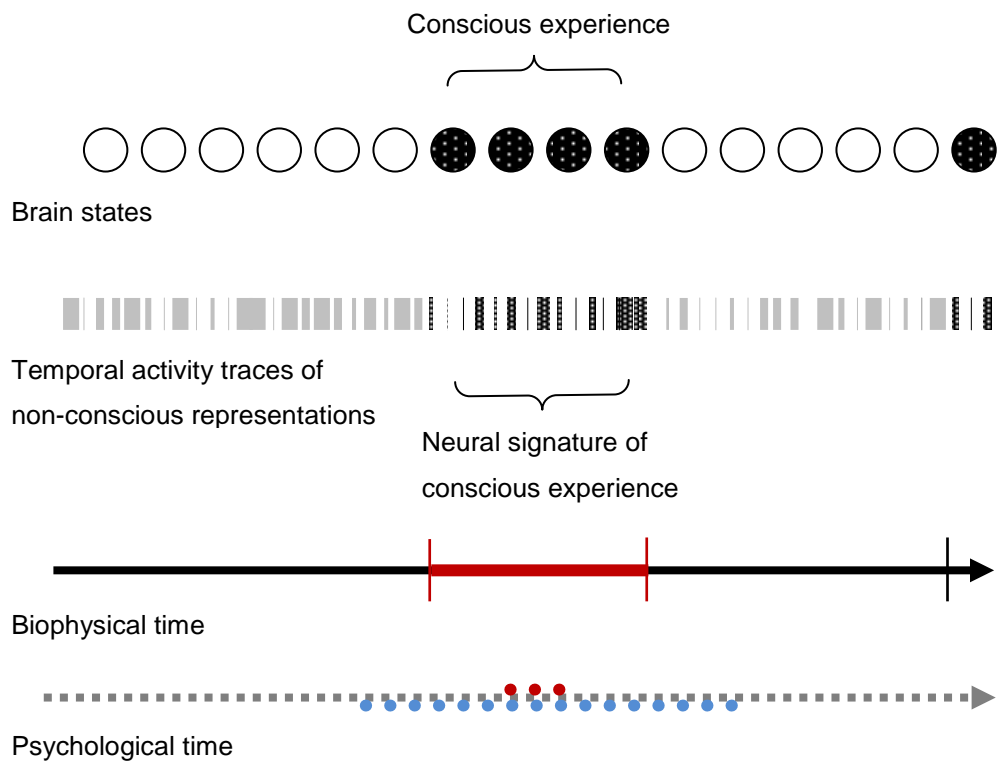

## 5. Conclusions

Non-conscious brain representations are the basis upon which all conscious experience is built. The capacity of the human brain to generate such experience is a result of evolution, expressed through continuous interaction between the brain and its environment from early childhood on, progressively enabling conscious experience on the basis of the temporal matching of the neural activity traces of non-conscious representations. These matching processes take place in physiologically determined biophysical time, while psychological time associated with a conscious experience is entirely subjective. The brain learning process which ensures the continuity and the stability of representations and therefore that of conscious experience relies, as suggested by the model approaches discussed here above, on the progressive development of dedicated resonant circuits capable of reentrant signaling and space-time signal de-correlation. The latter ensures that the temporal activity traces of non-conscious representations are maintained in the brain independently from functional specialization or spatial cortical maps. The dedicated resonant circuits that are necessary to achieve this are progressively and arbitrarily formed in the brain. Although their intrinsic functional properties are universal and pre-wired, their expression strongly depends on epigenetic factors. The latter determines the amount of long-range connectivity between neural structures activated by bottom-up signals at a given moment in time, and distant structures not directly activated by bottom-up input. When a critical amount of such long-range circuitry is consolidated, reentrant signaling will trigger conscious experience. This happens whenever non-conscious representational traces statistically match the temporal signatures of learnt and sufficiently stable long-term memory representations. Further investigation of experience related temporal activities in the thalamo-cortical and cortico-cortical pathways of the brain might one day allow tracing such mechanisms.